\documentclass{camera}
\usepackage{graphicx}  

\begin{document}

%
\title{Characterization of quasi-projectiles produced in symmetric collisions studied with INDRA \\ Comparison with models}

%
\author{E. Legou\'{e}e, E. Vient \and D. Durand}

%
\organization{LPC Caen, Universit\'{e}, ENSICAEN, CNRS/IN2P3, F-14050 Caen cedex, France}

\maketitle

\begin{abstract}

The characterization of hot quasi-projectiles produced in symmetric or quasi-symmetric reactions (Au + Au, Xe + Sn, Ni + Ni, Ar + KCl) at different incident energies are estimated by means of two different procedures. The advantages and disadvantages of each method are analyzed on the basis of simulations using events produced by two slightly different models: HIPSE and ELIE. 
\end{abstract}

%

\section{Introduction}
 
	Nuclear reactions around the Fermi energy produce hot species by means of energy and/or mass transfers between the two partners of the collision. Of particular interest is the production of hot projectile-like nuclei because, due to favourable kinematical conditions, they can be accurately detected by large multidetectors such as INDRA \cite{Pouthas}. The thermodynamics of such hot species is a key topic of nuclear physics since it allows to address, for instance, the existence of a liquid-gas phase transition in nuclear matter. The precise measurement of the amount of excitation energy deposited in hot nuclei (calorimetry) is a prerequisite in such studies. The complexity of nuclear reactions makes such measurements rather difficult and requires accurate and well-controlled methods. These latter use event selections and kinematical cuts based on the angular distributions of all detected charged particles. The aim of the present work is to discuss optimal procedures for nuclear calorimetry. For this, computer simulations modeling nuclear reactions can be helpfull. Here, we have used the events generated by two models: HIPSE (Heavy Ion Phase Space Exploration) \cite{Lacroix} and ELIE \cite{Durand_IWM07}.

\section{Brief description of the events generators and comparison with selected data}

	The two event generators are based on a two-step scenario of the reaction: 
\begin{itemize}
\item
an entrance channel phase ending with the formation (for finite impact parameters) of excited PLF (projectile-like fragment), TLF (target-like fragment) and participants. 
\item
a second phase considering secondary decay and propagation towards the detectors.  
\end{itemize}

In the ELIE model, the mass numbers of the PLF, TLF and participants are obtained by considering the geometrical overlap of the nuclei for each impact parameter according to the high energy participant-spectator picture while in the HIPSE case, it is governed by the minimum distance of approach obtained with a trajectory calculation based on a realistic interaction potential.

 To build the kinematics of the projectile-like, the target-like and the partition of the participants, the following 
hypothesis are assumed:
\begin{enumerate}
\item
The momentum distribution of the incoming
nucleons inside the two partners is supposed to have no time to relax on a time scale comparable with the
reaction time. This is a frozen approximation: only a few hard nucleon-nucleon collisions can occur and those latter are governed by a single parameter: the mean free path.

\item
In the ELIE model, the partition of the participants is generated by a random process in momentum space. The mass 
number A of each species (including A=1 free nucleons) is sequentially chosen at random by
picking A nucleons from the nucleon momentum distribution. For IMF's ($A \geq 4$), the excitation energy, $E^*$, is obtained by summing the center-of-mass kinetic energy of all nucleons belonging to the fragment. If $E^*$ is larger than a maximum value associated with a maximum temperature $T_{max}$, the fragment is rejected and a new try is made until all nucleons have been assigned. In the following, it turns out that a value of $T_{max}$ = 5.5 MeV allows to reproduce the experimental data (with a level density parameter equal to A/10, $E^*$= 3 MeV/u). 

\item
In the HIPSE approach, an aggregation algorithm based on cuts in momentum and real space is applied to build the fragments and constitute the partition.
\end{enumerate}

In a second step, the partition is propagated in space-time and secondary decays are considered. This is done using a standard evaporation code: the SIMON event generator \cite{Lauwe}. The essential difference between the two models lies in the excitation energy distribution in the fragments since in the ELIE case, the limiting temperature (5.5 MeV) limits E* to around 3 MeV/u while for HIPSE, the amount of excitation energy can extend to much larger values (see blue histograms in fig.\ref{fig03}) (colour online).

\begin{figure}
\centerline{\includegraphics[width=11cm,height=9cm]{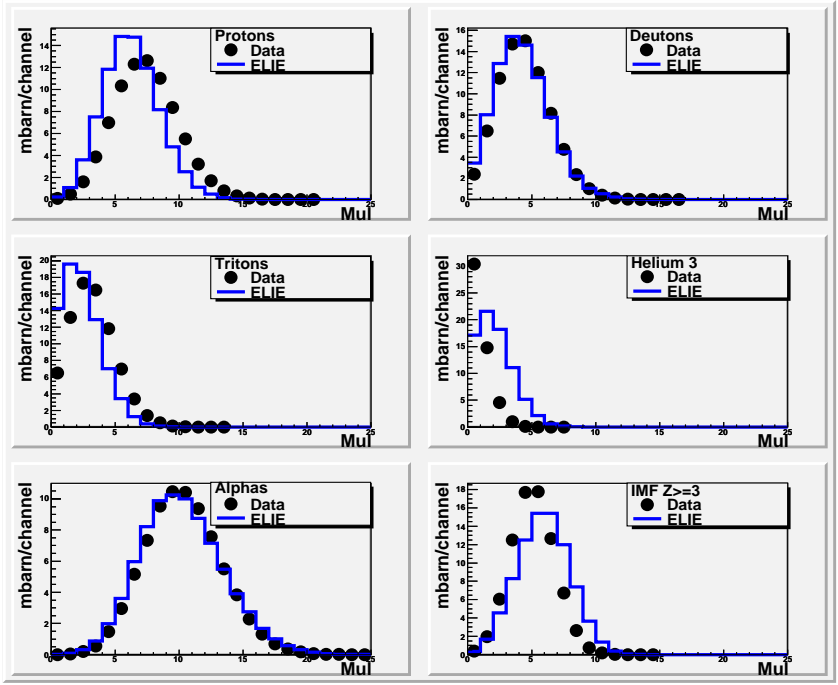}}
\caption{Particle multiplicities cross-sections(from protons to IMF as indicated in each panel) for $Xe$ + $Sn$ central collisions at 50 A.MeV. Black points are INDRA data while the histograms are for ELIE data (Colour online).}
\label{fig02} 
\end{figure}

	Figure \ref{fig02} shows a comparison of the multiplicity cross-sections from protons to IMF's between INDRA data and ELIE data for Xe+Sn central collisions at 50 A.MeV. For that purpose, a selection of the events has been made using the total light charged particle transverse energy, $E_{tr12}$, normalized to the CM available energy. Assuming that this latter is correlated to the impact parameter, central collisions have been defined by keeping only the 50 mbarn cross-section associated with the largest $E_{tr12}$ values. For such central collisions, particle multiplicities are correctly reproduced by the model. Other observables have been considered and a general good agreement has been reached between INDRA data and ELIE data. Same conclusions can be reached as far as HIPSE is concerned. These models are therefore sufficiently realistic to allow a check of the calorimetric methods described in the following.

\section{Data analysis and calorimetries}

  PLF's characteristics (mass, charge and excitation energy)  are measured by considering fragments and light charged particles on an event by event basis. The question of the undetected neutrons is discussed later. A completion criterium is first applied to the data by imposing that most of the charged particles (around 80 percent at the forward of the center of mass) emitted in the reaction be detected. This ensures that no major biases are introduced in the final retained event sample. Then, two different methods are used: a standard calorimetry and the so-called 3D calorimetry. 
	The projectile-like frame \cite{Durand} in which all kinematical quantities are calculated is built by adding all fragments (charge larger than 2) which are located in the forward hemisphere of the center of mass of the reaction (for symetric systems this corresponds to positive CM velocities).
	Considering the standard calorimetry, all particles emitted in the forward hemisphere of the PLF frame are assumed to be emitted by the PLF and as such are taken into account in the reconstruction of the excitation energy and mass. In order to cover the whole 4$\pi$-space, this contribution is doubled on an event by event basis. Note that in this case, no attempt is made to consider non-equilibrium effects. 

\begin{figure}
\includegraphics[width=12cm,height=8cm]{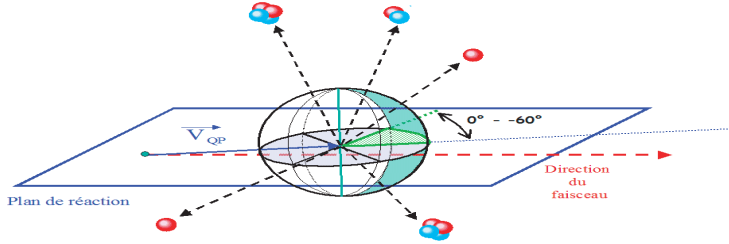}
\caption{Angular domains with respect to the reaction plane and beam direction chosen to determine the probability of emission of particles by the PLF.}
\label{fig01} 
\end{figure}

	Such effects are tentitatively taken into account in the 3D calorimetry by first assuming that PLF evaporation is 
cylindrically symmetric around the perpendicular axis of the reaction plane, as shown in fig.\ref{fig01}. It is then assumed that all particles emitted in the angular domain  $\varphi$ [{$0^\circ$}, {$-60^\circ$}] result from the evaporation of the PLF, and, as such, the evaporation probability in this angular range $prob_{ref}(i)$ of species $i$ (proton, deuteron, triton, ...) is equal to 1 whatever the kinetic energy (${E_c}_i$) distribution, called $W_{ref}(i)$. In other angular regions covering the same solid angle, other processes of emission can occur so that the evaporation probability can take values lower than 1. The probability $p_j(i)$ in this region labelled $j$ is then obtained by dividing the kinetic energy distribution $W_j(i)$ by $W_{ref}(i)$ for each energy bin $E_c(i)$. Thus, with this method, the probability of evaporation is a function of the species (proton, deuteron, triton, ...), the kinetic energy and the angle of emission of the particle. The probability of evaporation hereafter labelled $prob_i$ for particle labelled $i$ is used for the reconstruction of the hot PLF : in mass $(A_{PLF})$, charge 
$(Z_{PLF})$, and excitation energy ${{E^*}_{PLF}}$ according to the following equations:
	
\begin{equation}
 {A_{PLF}} = {\sum_{i=1}^{multot} prob_i \times A_i + N_{neutron} }
\end{equation}

\begin{equation}
 {Z_{PLF}} = {\sum_{i=1}^{multot} prob_i \times Z_i }
\end{equation}

\begin{equation}
 {{E^*}_{PLF}} = {\sum_{i=1}^{multot} prob_i \times {E_c}_i + N_{neutron} \times {\langle E_c \rangle}_{p + \alpha} - \Delta Q}
\end{equation}

	The number of neutrons is obtained by imposing the conservation of the $(A/Z)$ of the projectile. The mass number of fragments when not measured is obtained from the measured atomic number following the Charity prescription \cite{Charity}.
\\	
 
\begin{figure}
\begin{tabular}{cc}
	\centerline{\includegraphics[width=6.2cm,height=6cm]{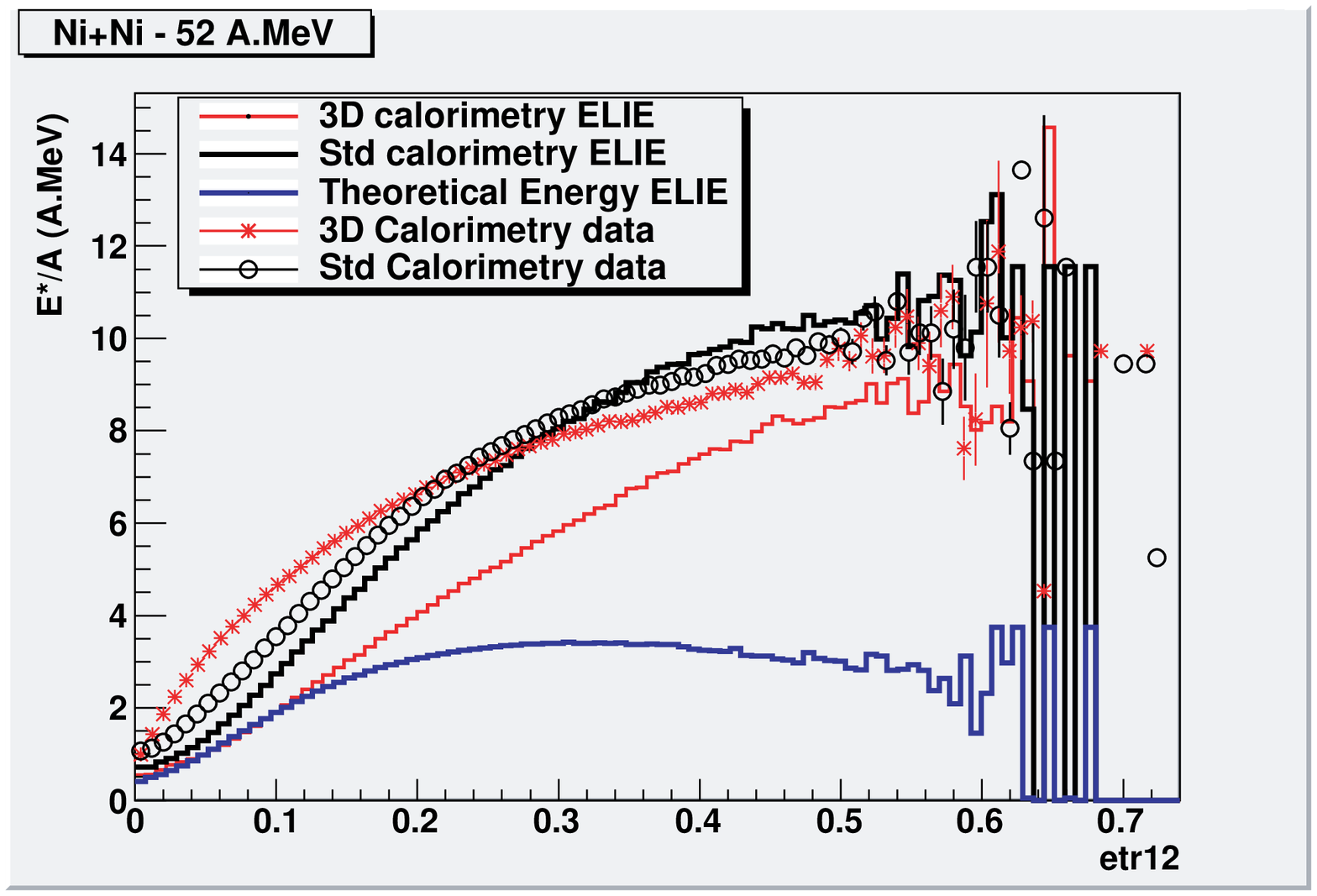}
	\includegraphics[width=6.2cm,height=6cm]{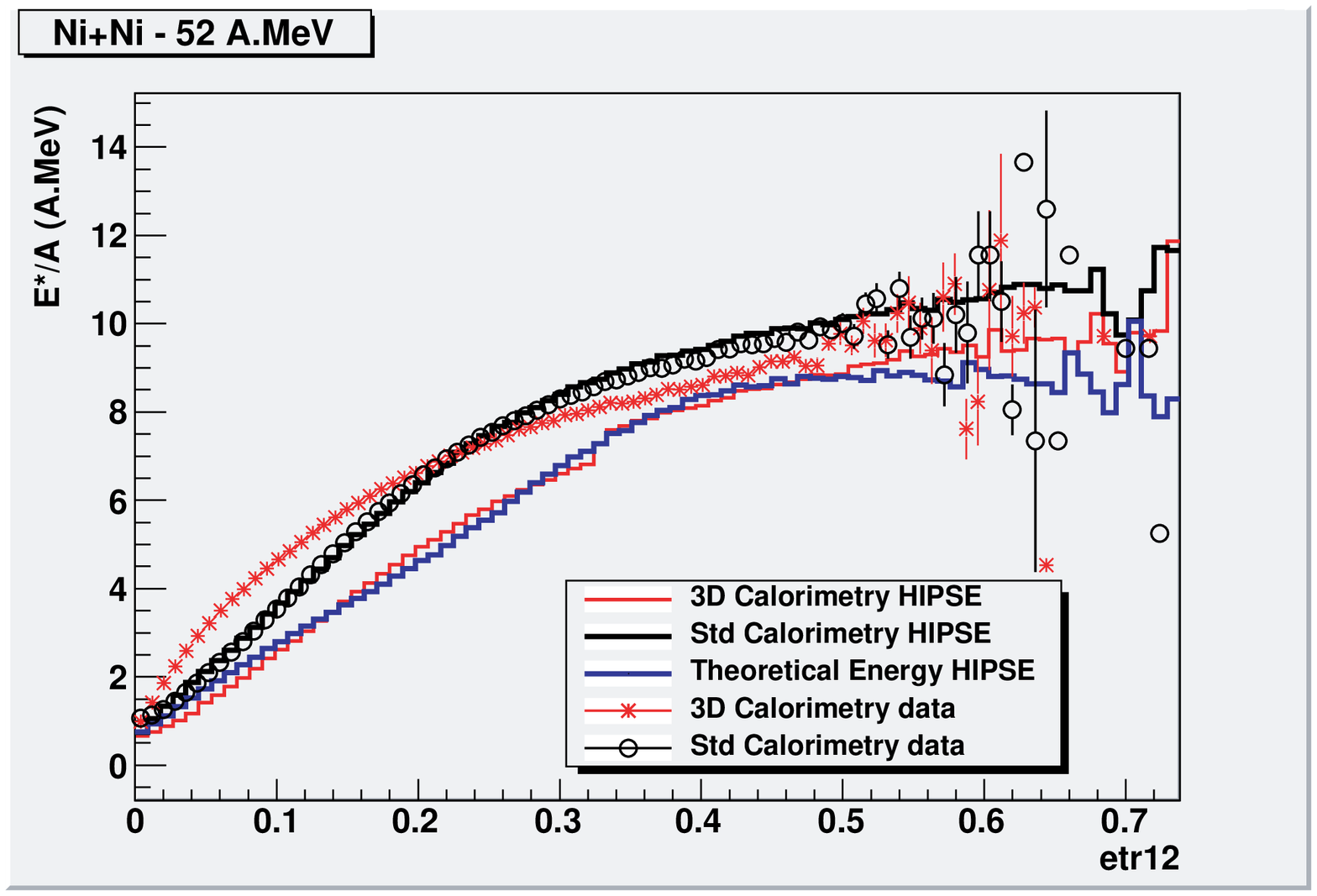}}
\end{tabular}
\caption{$<{{E^*}_{PLF}}>$ (MeV/u) as a function of 
$E_{tr12}$ for Ni+Ni at 52 A.MeV. Left, ELIE data. Right: HIPSE data, both compared with INDRA data (red crosses and open points)(Colour online).}
\label{fig03}
\end{figure}

	Figure \ref{fig03} shows preliminary results concerning the correlation between ${E}_{tr12}$ and the mean PLF excitation energy per nucleon $\frac{E^*}{A}$ for ELIE data (left) and HIPSE data (right). In both cases, the INDRA data have been superimposed for a sake of comparison. 
	 
	Results of the studies are rather contrasted. None of the methods can reproduce the excitation energy values given by the ELIE model and a strong over-estimation is obtained especially for the most central collisions. This is due to the fact that, in the ELIE case, even in the reference angular domain, most of emitted particles do not originate from a slow equilibrated process. However, for excitation energies lower than 2 MeV/u, the 3D method gives good results. As for HIPSE, 
the 3D calorimetry reproduces the excitation energy distribution while the standard method leads to a systematic over-estimation. It is worth noted that, whatever the method,  the methods applied to INDRA data lead to very high excitation energies: a point which remains to be understood.

\section{Summary and perspectives}

	Two different methods for the characterization of projectile-like fragments in nuclear reactions have been discussed. Tests of the procedures using data generated by two models, ELIE and HIPSE, have been shown. Results are contrasted. However, work is still in progress to improve the 3D calorimetry. This could help extract, by subtraction of the evaporation component, the non-equilibrium processes associated with the first instants of the reaction.




%
\end{document}